\documentclass[prb,aps,twocolumn,superscriptaddress,showpacs]{revtex4}

\usepackage{graphicx}

\begin{document}

\title{Central mode and spin confinement near the boundary of the superconducting phase in YBa$_{2}$Cu$_{3}$O$_{6.353}$ (T$_c$=18 K)}

\author{C. Stock}

\affiliation{Department of Physics, University of Toronto, Ontario, Canada M5S 1A7}

\affiliation{Department of Physics and Astronomy, Johns Hopkins University, Baltimore, MD USA 21218}

\author{W.J.L. Buyers}

\affiliation{National Research Council, Chalk River, Ontario, Canada K0J 1J0}

\affiliation{Canadian Institute of Advanced Research, Toronto, Ontario, Canada M5G 1Z8}

\author{Z. Yamani}

\affiliation{National Research Council, Chalk River, Ontario, Canada K0J 1J0}

\author{C.L. Broholm}

\affiliation{Department of Physics and Astronomy, Johns Hopkins University, Baltimore, MD USA 21218}

\author{J.-H. Chung}

\affiliation{NIST Center for Neutron Research, Gaithersburg, Maryland USA 20899}

\affiliation{Dept. of Materials Science and Engineering, University of Maryland, College Park, MD 20742, USA}

\author{Z. Tun}

\affiliation{National Research Council, Chalk River, Ontario, Canada K0J 1J0}

\author{R. Liang}

\affiliation{Physics Department, University of British Columbia, Vancouver, B. C., Canada V6T 2E7}

\affiliation{Canadian Institute of Advanced Research, Toronto, Ontario, Canada M5G 1Z8}

\author{D. Bonn}

\affiliation{Physics Department, University of British Columbia, Vancouver, B. C., Canada V6T 2E7}

\affiliation{Canadian Institute of Advanced Research, Toronto, Ontario, Canada M5G 1Z8}

\author{W.N. Hardy}

\affiliation{Physics Department, University of British Columbia, Vancouver, B. C., Canada V6T 2E7}

\affiliation{Canadian Institute of Advanced Research, Toronto, Ontario, Canada M5G 1Z8}

\author{R.J. Birgeneau}

\affiliation{Department of Physics, University of California at Berkeley, Berkeley, CA 94720}

\affiliation{Canadian Institute of Advanced Research, Toronto, Ontario, Canada M5G 1Z8}

\date{\today}

\begin{abstract}

We have mapped the neutron scattering spin spectrum at low-energies in YBa$_{2}$Cu$_{3}$O$_{6.353}$ (T$_{c}$=18 K) where the doping $\sim$0.06 is near the critical value ($p_{c}$=0.055) for superconductivity. No coexistence with long range ordered antiferromagnetism is found.  The spins fluctuate on two energy scales, one a damped spin response with a $\simeq$2 meV relaxation rate and the other a central mode with a relaxation rate that slows to less than 0.08 meV below T$_{c}$. The spectrum mirrors that of a soft mode driving a central mode.  Extremely short correlation lengths, 42$\pm$5 \AA\ in-plane and 8$\pm$2 \AA\ along the c direction, and isotropic spin orientations for the central mode indicate that the correlations are subcritical with respect to any second order transition to N\'{e}el order.  The dynamics follows a model where damped spin fluctuations are coupled to the slow fluctuations of regions with correlations shortened by the hole doping.

\end{abstract}


\maketitle

Compared to the extensive studies of spin fluctuations deep within the superconducting (SC) phase of cuprates,~\cite{Fong00:61,Dai01:64} less is known for low doping. There is little consensus as to the nature of the precursor phase from which the SC phase emerges when holes are doped into a planar array of correlated S=$1/2$ Cu spins.  Whereas superconductivity in La$_{2-x}$Sr$_{x}$CuO$_{4}$ (LSCO) emerges from a non-metallic phase, the precursor phase for YBa$_{2}$Cu$_{3}$O$_{6+x}$ (YBCO$_{6+x}$) remains controversial~\cite{Lavrov98:41,Loram01:62,Sutherland05:94,Sun-Ando05:72}. Probing the spin structure in the vicinity of the boundaries of the antiferromagnetic (AF) and SC phases is necessary to determine whether 
antiferromagnetism and superconductivity compete, coexist, or are separated by a novel phase.

For single layer LSCO, an insulating phase with frozen short-range magnetic order separates the AF ordered from the SC phase~\cite{Kastner98:70,Keimer92:46,Matsuda02:65,Wakimoto00:62}. At the SC boundary, the magnetic structure abruptly changes by a 45$^\circ$ rotation of the incommensurate wave vector of static spin stripes, so the transition to superconductivity is first order. In contrast, the temperature dependence of the commensurate scattering in La$_{2}$CuO$_{4}$ lightly doped with Li is interpreted as indicating the proximity of a quantum critical point (QCP)~\cite{Bao03:91}. It is clearly important to study other cuprates such as YBCO$_{6+x}$ to determine the nature of the state near the edge of the SC phase.

For the bilayer system YBCO$_{6+x}$ oxygen doping occurs in the chains well away from the planes and complex structural changes as in LSCO are not evident. There have been suggestions that the AF and SC phases may meet at a point~\cite{Shirane90:41,Ando99:83}.  Recently Sanna \textit{et al.}~\cite{Sanna04:93} claimed from $\mu$SR data that the ordered AF state coexisted with superconductivity with a N\'{e}el temperature exceeding that of a freezing transition to a `cluster spin-glass' state. This contrasts with LSCO and Y$_{1-y}$Ca$_{y}$Ba$_{2}$Cu$_{3}$O$_{6+x}$ where AF and SC phases are well separated as shown in Fig.4 of Sanna.~\cite{Niedermayer98:80}  

Another unique aspect of YBCO$_{6+x}$ is that the SC phase reveals a resonance peak whose energy scales with T$_{c}$ for at least a limited range of doping~\cite{Fong00:61,Dai01:64,Stock04:69}.  Quantum critical theories suggest that the resonance peak may be a soft mode of the SC phase~\cite{Chubukov94:49} and therefore it is crucial to study spin behavior close to the critical doping for superconductivity of p$_{c}$=0.055~\cite{Matsuda02:65} to test whether the antiferromagnetic QCP lies within the SC phase or whether it is contiguous.

Here we approach the SC boundary by studying YBCO$_{6.353}$ where T$_{c}$=18 K is suppressed to one fifth of optimal.  The hole density of p$\sim$0.06 exceeds the critical value ($p$=0.055) by only $\sim$9$\%$. Earlier results~\cite{Mook02:88} with a similar stated oxygen concentration gave a much higher T$_{c}$ and so lie much farther from the critical doping.  We will show that just beyond the critical doping the spins fluctuate on two distinct time scales, one giving a broad energy feature and the other a central mode at zero energy.  A single spectral form is found in which damped spin fluctuations transfer strength to an intense central mode with an extremely narrow energy width similar to the spectrum near antiferroelectric transitions~\cite{Shapiro72:6,Halperin76:14}.   The absence of long-range spin order and short spatial correlations indicate that the system may lie farther from a second order phase transition to long-range AF order than the lower critical concentration for the SC phase. Taken with the conservation of spin orientational symmetry found with polarized neutrons, the spins display, not a freezing transition to a cluster phase, but a gradual formation of only short-range correlations that are intrinsic to the SC phase.  

Thermal neutron measurements were made with the DUALSPEC spectrometer at Chalk River Laboratories using PG(002) as monochromator and analyzer and a PG filter in the scattered beam.  The horizontal collimation was set to [33$'$ 29$'$ \textit{S} 51$'$ 120$'$] and [33$'$ 48$'$ \textit{S} 51$'$ 120$'$] for E$_{f}$=14.6 meV and 5 meV, respectively.  The setup for polarized beam measurements was described previously.~\cite{Stock04:69}  Cold-neutron measurements were made with the SPINS spectrometer at the NCNR with a PG (002) monochromator and analyzer. For E$_{f}$=2.9 meV the collimation was [guide 80$'$ \textit{S} 80$'$ 80$'$].  For energy transfer less than 2 meV a Be filter was placed before the monochromator and above 2 meV a BeO filter was placed in the scattered beam. 


\begin{figure}[t]

\begin{center}

\resizebox{1.0\linewidth}{!}{\includegraphics{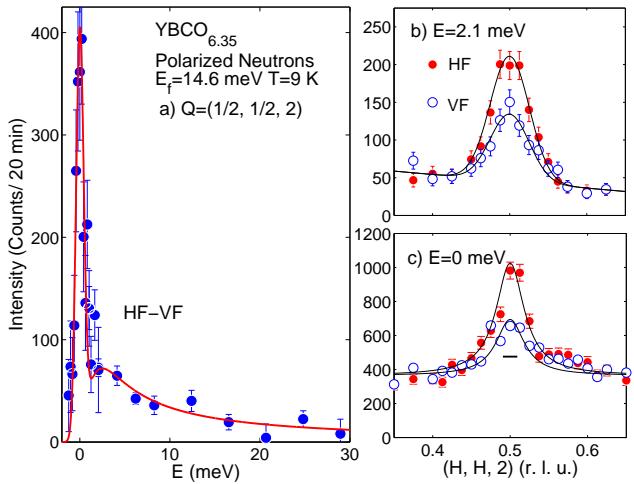}}

\end{center}

\caption{\label{spectrum}  (Color online) Polarized spin-flip neutron scattering.  \textit{a}) The magnetic spectrum along [1,$\overline{1}$,0] measured by subtracting the VF from the HF data. Both inelastic \textit{b}) and elastic scans \textit{c}) show the spin directions are isotropic since the HF intensity is twice that for VF.  The horizontal bar in \textit{c} represents the resolution full-width.}

\end{figure}

\begin{figure}[t]
\begin{center}
\resizebox{0.9\linewidth}{!}{\includegraphics{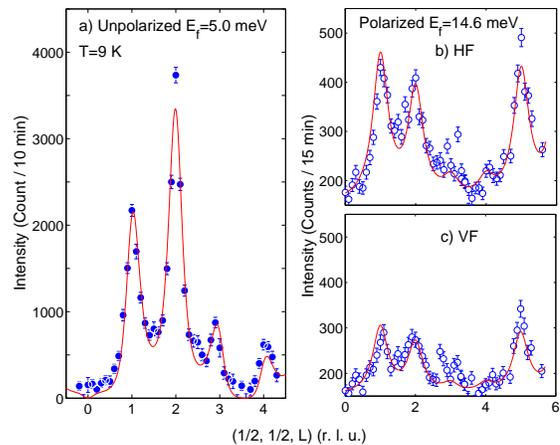}}
\end{center}

\caption{\label{polarised}  (Color online) Elastic scans \textit{a}) Comparison of short-range incipient 3D correlations normal to the planes and the coupled bilayer model (line). An 80 K background has been subtracted.  \textit{b}) and \textit{c}) Polarized neutron spin-flip scattering for horizontal and vertical field.  The fact that HF=2VF independent of L shows that the scattering is isotropic. The enhancement of the L=1 peak for E$_{f}$=14.6 meV relative to 5 meV arises from resolution effects.}

\end{figure}

Seven $\sim$1 cc crystals were co-aligned in the (H,H,L) scattering plane with an overall mosaic of $\sim$1.5$^{\circ}$. From DC magnetization (Fig. 3c) the critical temperature was measured to be T$_{c}$=18 K with a width of $\sim$2 K.   The orthorhombic lattice constants were $a$=3.843, $b$=3.871, and $c$=11.788 \AA. Superlattice peaks associated with CuO chain ordering were not observable indicating that oxygen ordering in the chains is short-ranged.

Polarized neutron measurements of the HF-VF subtracted spin-flip scattering (Fig. \ref{spectrum}a) probe the spin fluctuation spectrum along the [1,$\overline{1}$,0] direction.  They show that the spins fluctuate on two energy scales, one showing a resolution limited central peak ($\hbar \omega$=0) with very slow dynamics, and another with a faster timescale giving a broad inelastic peak at $\sim$2 meV (see also Fig. \ref{central}). 

Scans through the AF wave vector along [H,H,2] (Fig. \ref{spectrum}b, c) and along [$\frac{1}{2}$,$\frac{1}{2}$,L]  (Fig. \ref{polarised}b, c) show that the HF intensity is twice the VF intensity.  It follows that the fluctuations are equally strong within and normal to the (H,H,L) scattering plane, even as the varying L changes the amount of [0,0,1] fluctuations sampled. This shows that the spins are isotropic in direction for the near-static central mode.  Fig. \ref{spectrum}b shows that inelastic fluctuations for  {\bf{Q}}=($\frac{1}{2}$,$\frac{1}{2}$,2) are also isotropic. At all temperatures the spin rotational symmetry is conserved. This state clearly differs from the anisotropic critical fluctuations that would have been seen were the system approaching an AF phase, in which moments aligned in-plane would have caused an energy gap for out-of-plane fluctuations polarized along [001].  The low-temperature isotropic magnetic ground state observed in YBCO$_{6.353}$ is clearly dramatically different from that of the insulator.  The static spin cross-section as shown later is dynamically coupled to spin excitations and so is not the result of a phase-segregation of holes or a doping gradient.  If this had been so, the clusters would have mimicked the properties and anisotropy of the antiferromagnetic insulator.  It is presently not clear how this ground state relates to that suggested to explain NMR results in LaCuO$_{4+\delta}$, crystal field excitations in ErBa$_{2}$Cu$_{3}$O$_{x}$, and $\mu$SR data in LSCO and Ca doped YBCO.~\cite{Hammel90:42,Mesot93:70,Niedermayer98:80}
 
From scans  with E$_{f}$=2.9 meV along [110] we extract an in-plane correlation length of 42$\pm$5 \AA\ for the central mode from a resolution-convolved fit to a lorentzian. 

The integer-centered peaks along L in Fig. \ref{polarised}a reflect the lattice periodicity, but have a large width.  The elastic scattering is well described by a model where internally AF bilayers are weakly coupled along [001]. The structure factor along L for fixed in-plane wave vector ($\frac{1}{2}$,$\frac{1}{2}$) is proportional to

\begin{eqnarray}
\label{gap_acoustic} {F^{2}(L) = {{\sin^2(\pi L d/c)\sinh^{2}(c/\xi)}\over{(\cosh(c/\xi)-\cos(2\pi L))^{2}}}}
\end{eqnarray}

\noindent with $\xi$ the correlation length and $d$ the bilayer separation ($d/c$=0.28), and to the anisotropic Cu$^{2+}$ form factor~\cite{Shamoto93:48}. The resolution convolved fit (solid line) gives a dynamic correlation length of only 8$\pm$2 \AA\ along [001].   This interplanar correlation of less than one unit cell shows that the spins are primarily correlated within the two-dimensional (2D) CuO$_{2}$ bilayers. The short-range nature of the spin correlations indicate that YBCO$_{6.353}$ is strongly sub-critical with respect to 3D long range AF order even at low temperature.   The presence of short-ranged elastic correlations and two distinct timescales differs substantially from the behavior near a second-order phase transition where the energy linewidth is directly related to and scales with the correlation length. We emphasize that short-ranged spin correlations do not imply the formation of physical clusters except in the most naive interpretation; a uniform state exhibiting superconductivity and finite spin correlations is a possible ground state.

\begin{figure}[t]

\begin{center}
\resizebox{0.8\linewidth}{!}{\includegraphics{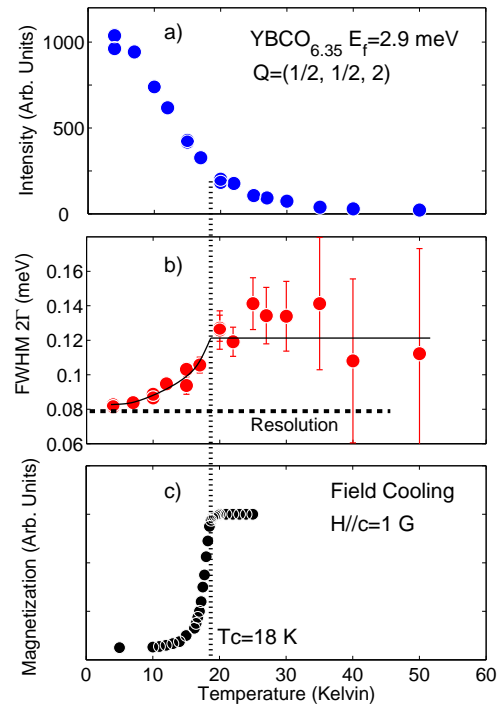}}
\end{center}

\caption{\label{quasi}  (Color online) Panel a) illustrates the
growth on cooling of the peak intensity of the central mode from a Gaussian fit and compares it with the volume fraction of static spins measured by Sanna \textit{et al.} using muons. Panel b) shows how the full-width of the central peak drops smoothly to the resolution limit (broken line) below T$_{c}$ (dotted line) and panel c) shows the sharp Meissner transition at 18 K.}

\end{figure}


Fig. \ref{quasi} displays the peak intensity and energy width of the central mode spectrum from Gaussian fits.  The peak intensity grows on cooling, dramatically and smoothly (panel a), with no sign of a transition, and is unperturbed by the sharp SC transition at 18 K (panel c).   On cooling, the central mode energy width declines to the resolution limit of $\sim$0.08 meV (panel b). A possible picture is of spins coupled through a fermionic response associated with holes. 


\begin{figure}[t]
\begin{center}
\resizebox{0.8\linewidth}{!}{\includegraphics{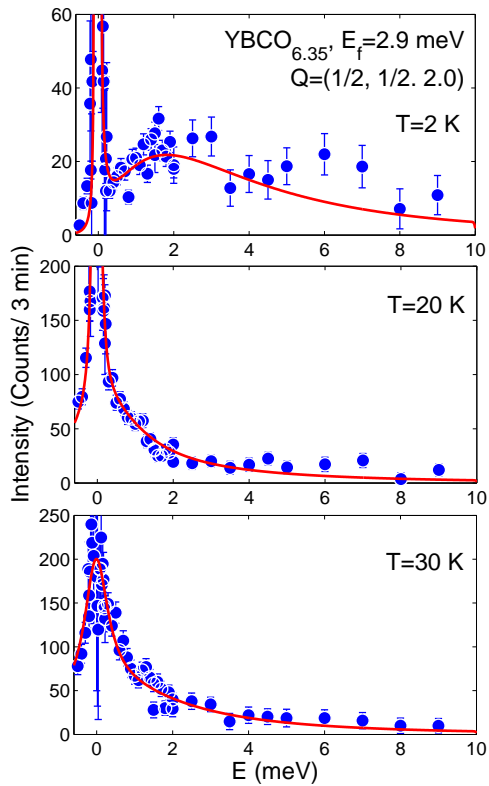}}
\end{center}

\caption{\label{central}  (Color online) Temperature evolution of the background subtracted spin spectrum at {\bf{Q}}=($\pi$,$\pi$).  The line shows the response of a damped mode that on cooling softens and drives up the intensity of a central peak arising from slow hopping of a spin cluster (see text).  The parameters for the fits at T=2, 20, 30 K respectively are $\hbar \Omega_{0}$=0.9 $\pm$ 0.3, 2.4 $\pm$ 1.1, 3.4 $\pm$ 1.9 meV; $\tau/\hbar$= 14.9 $\pm$ 3.0, 9.2 $\pm$ 1.2, 1.8 $\pm$ 0.8 meV$^{-1}$; $\hbar \delta$= 3.9 $\pm$1.3, 2.0 $\pm$0.9, 2.0 $\pm$0.9 meV.  }

\end{figure}

The appearance of a sharp central mode, whose energy scale is much smaller than that of a damped inelastic mode, bears a strong similarity to the soft and central mode spectrum in SrTiO$_{3}$.  Halperin and Varma~\cite{Halperin76:14} and Shapiro \textit{et al.}~\cite{Shapiro72:6} showed that a coupling of a damped harmonic mode to a slowly fluctuating object would give rise to a central mode. We surmise that the object in YBCO$_{6.353}$ results from the slow fluctuations of the many spins within a correlation range frustrated by the metallic hole doping.

The spectrum  at $(\pi,\pi)$ (Fig. \ref{central}) was obtained from the difference between constant-${\bf{Q}}$ scans at the peak of the magnetic scattering, {\bf{Q}}=($\frac{1}{2}$,$\frac{1}{2}$,2), and at a background position (0.35, 0.35, 2).  The solid lines are the result of fits to the phenomenological model~\cite{Halperin76:14,Shapiro72:6}

\begin{eqnarray}
\label{spectral_form} {S(\omega) \propto
[n(\omega)+1]\textrm{Im}\biggl(\Omega_{0}^2+\delta^2-\omega^2- i\omega \Gamma-{{\delta^2}\over {1-i\omega \tau}}\biggr)^{-1}}
\end{eqnarray}

\noindent which describes a mode of renormalized frequency $\Omega_{0}$ and damping $\Gamma$ coupled to an object that is hopping slowly at a rate $\gamma$=$1/\tau$.  The Bose factor is $[n(\omega)+1]$ and $\delta$ is a measure of the coupling between slow and fast spin modes.  The rapid growth of the central mode intensity on cooling (Fig. \ref{quasi}a) then arises, for fixed $\hbar \Gamma$=10 meV and amplitude,  from a softening of the harmonic mode $\Omega_{0}$ and from a slowing of the hopping rate $\tau^{-1}$.  The model of a soft mode driving a central-mode gives a good account of the two energy scales in Fig. \ref{central} and describes the data over a broad range of temperature and energy.  


In addition to the central mode, which is absent in YBCO$_{6.5}$~\cite{Stock04:69}, the spectrum of Figs. \ref{spectrum} and \ref{central} shows that the spin energy scale, defined by the $\sim$2 meV peak in $S({\bf{Q}},\omega)$, has decreased by an order of magnitude in YBCO$_{6.353}$ (Fig. \ref{spectrum}) from the 33 meV resonance energy~\cite{Stock04:69,Stock05:71} of YBCO$_{6.5}$.  This is larger than the doping ratio 0.06:0.09 and may indicate the predicted collapse of the spin energy on approach to the SC boundary~\cite{Chubukov94:49}.

In recent $\mu$SR measurements, Sanna \textit{et al.}~\cite{Sanna04:93} inferred that a freezing transition at T$_{f}$ to a `cluster spin-glass' state occurred within the SC phase. A direct comparison with the neutron results (see Fig. \ref{quasi}) shows different onset temperatures for the different probes.  This illustrates that T$_{f}$ (measured with muons) represents a slowing down beyond the longest muon timescale.  The freezing temperature does not represent a sharp transition.  The muon phase diagram~\cite{Sanna04:93} also suggests a coexistence of the SC phase and N\'{e}el order.  Our results prove there is no coexistence of long-range AF spin order just inside the SC dome.  Instead neutrons show, directly, that spin correlations are short-ranged and isotropic and that there is no phase transition to a 'glass' or 'cluster' state.  

We believe the slowing down of $\mu$SR in YBCO arises from  the same central peak as observed here in superconducting low-doped YBCO. 

The different temperatures where the slow dynamics becomes observable (Fig \ref{quasi}) reflect the timescales of the probes.  These results do not require the existence of literal clusters or of phase separation of holes.  Unlike neutrons, $\mu$SR cannot directly detect both spatial and temporal information of the spin correlations, but comparison shows that the muon data is consistent with the presence of the two energy scales and short-ranged correlations observed here. The short-range correlations coexist with the nascent superconducting ground state of YBCO - a complex ground state with slow fluid-like dynamics.~\cite{HermeleFisher05:72,LeeNagaosaWen06:78}


We have shown that in the superconducting state of YBCO a central mode arises from fluctuations that grow dramatically and become extremely slow in the low-temperature limit in the presence of  faster scale of relaxing spin excitations. The short-range correlations within and between planes show the ground state consists of sub-critical 3D-enhanced spin correlations. The isotropic spin orientations suggest that the spin configuration is radically changed from the antiferromagnet by the relatively small hole doping.

The NSF supported work at JHU through DMR 0306940 and work at SPINS through DMR-9986442.  Work at the University of Toronto and JHU was supported by NSERC.

\end{document}